\documentclass[prl,twocolumn,floatfix]{revtex4-1}
\usepackage{amssymb}
\usepackage{graphicx}
\usepackage{caption}
\usepackage{subcaption}
\usepackage{multirow}
\usepackage[table,xcdraw]{xcolor}
\usepackage{soul}
\usepackage{kantlipsum}
\usepackage{amsmath}
\usepackage{amsfonts}
\usepackage{amssymb}
\usepackage{dcolumn} 
\usepackage{braket}
\usepackage{booktabs}
\usepackage{multirow}
\usepackage[table]{xcolor}
\usepackage{dsfont}
\usepackage{times}
\usepackage{epsfig}
\usepackage{xcolor}
\usepackage{tikz}
\usepackage{float}
\usepackage{color}

\newcolumntype{P}[1]{>{\centering\arraybackslash}p{#1}}
\newcommand{\centered}[1]{\begin{tabular}{l} #1 \end{tabular}}

\def\bx {\bf x}

\graphicspath{ {Figures/} }

\usepackage{xr}
\makeatletter
\newcommand*{\addFileDependency}[1]{
  \typeout{(#1)}
  \@addtofilelist{#1}
  \IfFileExists{#1}{}{\typeout{No file #1.}}
}

\usepackage[english]{babel}

\begin{document}
\title[]{Flat Lensing by Graded Line Meta--arrays}

\author{Gregory~J.~Chaplain$^{1}$ and Richard~V.~Craster$^{1}$}
\affiliation{$^1$ Department of Mathematics, Imperial College London, London SW7 2AZ, UK }

\begin{abstract}
Motivated by flat lensing effects, now commonplace utilising negative refraction with finite slabs, we create negative refraction upon a graded line-array. 
We do so in the setting of flexural waves on a structured Kirchhoff--Love elastic plate, as a paradigm in wave physics, that has direct extensions to electromagnetic and acoustic wave systems. 
These graded line arrays are geometrically simple and provide strong coupling from the array into the bulk. 
Thorough analysis of the dispersion curves, and associated mode structure, of these meta--arrays, supported by 
 mode coupling theory, creates array guided wave (AGW) reversal and hybridisation into the bulk that leads to striking wave control via generalised flat lensing. 
 
\end{abstract}
\maketitle

 One of the dominant recent themes in wave physics, that has acted to re-energise the field, is that of using media, often based around metamaterials, that exhibit negative refraction. These ideas first gained traction in optics and electromagnetism \cite{pendry99a,smith04a}, and have also seen successful implementation across a broad range of wave systems such as in acoustics and elasticity \cite{liu00a,craster12a}, and are now seen as a paradigm for wave control more generally across physics. 

An attractive feature of such exotic, structured, negatively refracting materials is the ability to focus energy at a desired location as exemplified by the realisation of Veselago's \cite{Veselago1968} perfect flat lens in \cite{Pendry2000};  
the flat lens being a finite thickness slab of negatively refracting material. A variety of approaches, some as manipulations of photonic crystals, such as the rotation of the crystal used in all-angle negative refraction \cite{Luo2002,Lu2007}, have been used to induce flat lensing and this has been extended to three-dimensions and phononic crystals \cite{page04a,dubois19a} and latterly to elastic systems \cite{smith2012negative,brule14a, Dubois2013}. 

\begin{figure}[t!]
    \centering
    \includegraphics[width = 0.5\textwidth]{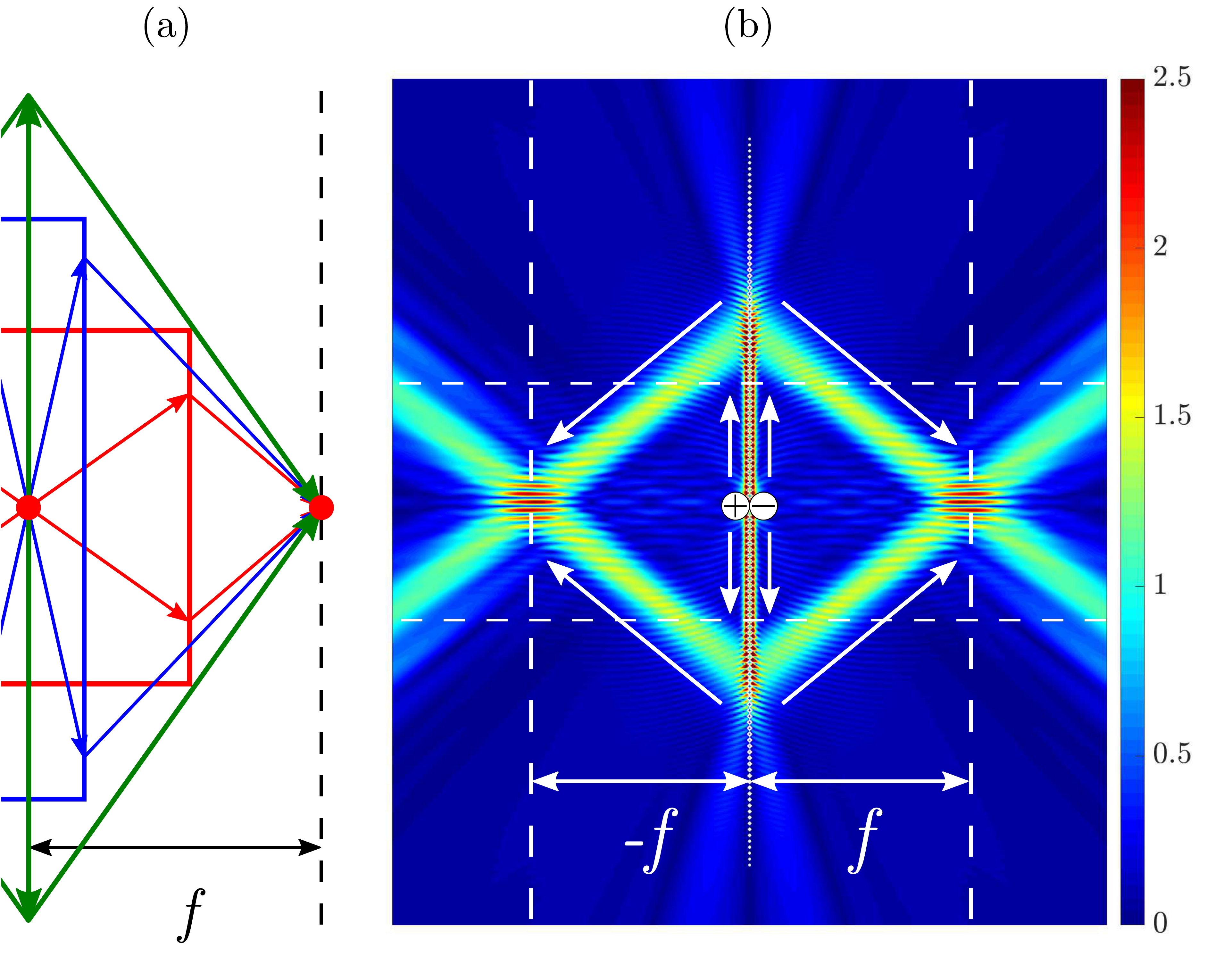}
    \caption{\textit{Flat lensing:}  (a)  Veselago-Pendry lenses decreasing in width until they form a line, each lens imaging a source at the central point to the same focal point. (b) Absolute wave-field for a graded meta--array, creating effective negative refraction on a line array, excited by a dipole source with polarity and position shown by white circles. White arrows show the path of the waves corresponding to the (green) rays of (a). Array characterised by $a = 1$ with point masses of $M = 10$. The array is graded beyond the horizontal dashed lines with grading index $\Delta r = -0.01$. All colourbars shown throughout are in arbitrary units.}
    \label{fig:FlatLens}
\end{figure}

This has inspired striking advances in the field of flat optics creating lensing effects at vastly reduced thicknesses compared to conventional optical components \cite{Yu2014,lalanne2017metalenses}; these are designed for the transmission of waves across structured interfaces whose subwavelength structurings introduce abrupt phase changes to the wavefield \cite{Fan2010, Aieta2009}. Structuring the interface, in optics and electromagnetism has also led to advances in leaky wave antennas as so-called holographic metasurfaces \cite{Fong2010,Minatti2011}. 
 We will take advantage of so-called rainbow trapping \cite{Tsakmakidis2007}, and extend it; in essence one can achieve a spatial separation of the spectrum by adiabatically grading an array through resonant or geometrical changes. Such trapping is used in electromagnetic settings \cite{Gan2009}, in acoustics \cite{Romero-Garcia2013} and recently in water waves \cite{bennetts18a} aiming at energy harvesting applications. In effect the grading of the array alters the phase and leads to wave to slow down, eventually to become trapped at a turning point and then reverse; different frequencies having different turning points. 

In this Letter, inspired by this success in optics and electromagnetism, we turn to grading arrays in elasticity. We design the structuring and grading of the array to induce phase changes, first by using rainbow trapping to reverse an array guided wave (and induce a phase shift upon reflection) and then by hybridising this reversed array wave into a beam in the bulk. By doing so, we obtain striking, and tunable, flat lensing as in Fig. \ref{fig:FlatLens}. These graded arrays possess much simpler spatial imhomogeneities than conventional phase-based flat lenses, and provide stronger coupling than slab-based flat lenses. This is shown in the context of flexural waves on thin Kirchhoff--Love (KL) elastic plates, although the analysis translates over to other wave regimes, providing motivation for new types of flat lens devices. We image a dipole source placed \textit{on} the surface of the graded array that acts as a line lens. Contrary to, say, the all-angle negative refraction flat lenses of \cite{Luo2002} the proposed graded meta--array flat lenses have strong coupling between source and device, as array guided waves are manipulated on the array itself, and thus the focal spot position can be tuned, via the grading profile, to lie many wavelengths away from the source as seen in Fig. \ref{fig:FlatLens}(b). 

Although there has been recent work on ungraded line arrays and their properties, 
\cite{haslinger16a,packo19a}, in flexural elastic wave systems there has been no attempt to grade them and explore this aspect. Thin plate flexural wave theories, \cite{landau70a,graff75a}, are widely used physical models for elastic waves in plates and provide  reliable predictions of many wave phenomena confirmed experimentally i.e. self-collimation \cite{lefebvre17a}, behaviour associated with Dirac cones \cite{torrent13a}, transformation optics for cloaking \cite{farhat09a}, negative refraction \cite{farhat10a} and more recently valley-hall edge states \cite{pal17a,makwana18a}. These flexural plate models also have a wider role in elasticity as motivation for the full vector elastic metasurface where rainbow-trapping by grading and mode-conversion from Rayleigh surface waves to bulk waves has been achieved using mode-conversion from the properties of the vector system \cite{colombi16a}.

\begin{figure}
    \centering
    \includegraphics[width = 0.45\textwidth]{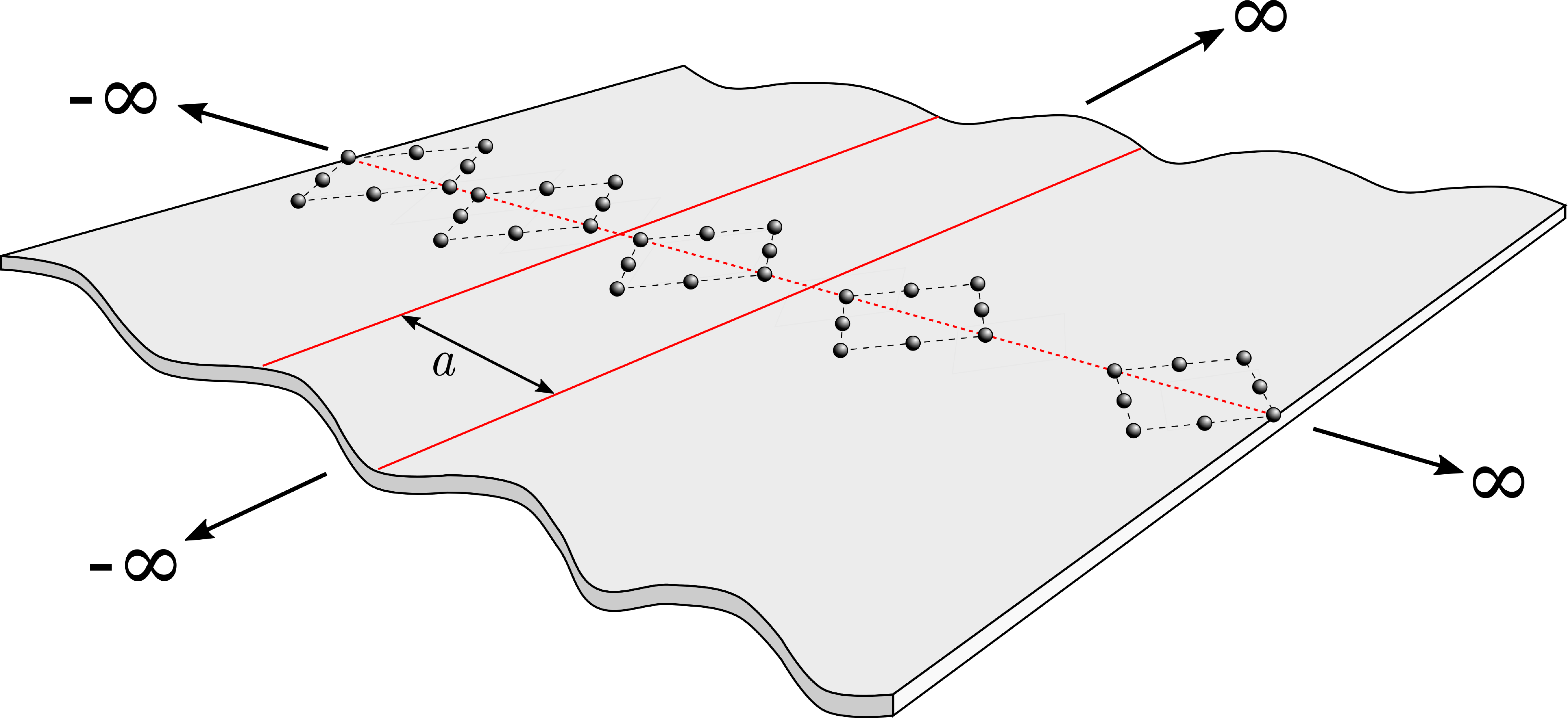}
    \caption{\textit{Graded diamond line array:} Elastic KL plate with graded array of point masses, with the line of symmetry/unit strip as dashed/solid red lines, respectively.}
    \label{fig:schematic}
\end{figure}

We choose to use the Kirchhoff-Love (KL) flexural wave model associated with wave propagation along a thin elastic  plate \cite{landau70a,graff75a} in order to take advantage of the explicit results for point scatterers that are available \cite{evans2007penetration}. A particularly attractive feature of the
KL model is that the fundamental Green's function is, unlike acoustics and electromagnetism, non-singular and thereby bounded which simplifies numerical simulations. The flexural waves on an infinite thin elastic plate, with
point constraints or forcing, are modelled using the vertical displacement field, $w({\bf x})$, satisfying the scalar (non-dimensionalised) KL equation
\begin{equation}
\left[\nabla^4 
 -\Omega^2\right]w({\bx})= F({\bf x}),
\label{eq:kirchhoff}
\end{equation} 
where the non-dimensionalised frequency
$\Omega^2=12(1-\nu^2)(\rho\omega^2)/(Eh^2)$, with $\rho$, $h$, $E$ and $\nu$ being the density, thickness, Young's modulus, and Poisson's ratio of the plate. The reaction forces, $F({\bf x})$, at point constraints introduce the dependence upon the geometrical arrangement of scatterers. 
One of the simplest models is that of a mass-loaded elastic plate that has the reaction forces proportional to the displacement and, for $P$ masses at positions ${\bf x}^{(p)}$,
\begin{equation}
F({\bf x})=\Omega^2\sum_{p=1}^{P}
 M^{(p)} w(\bx)\delta\left({\bf x}-{\bf x}^{(p)}\right).
\label{eq:ML_RHS}
\end{equation} 
We also use mass-spring loading, to take advantage of resonance, and we discuss that model in the Supp. Mat. 

\begin{figure}[b]
\includegraphics[width = 0.425\textwidth]{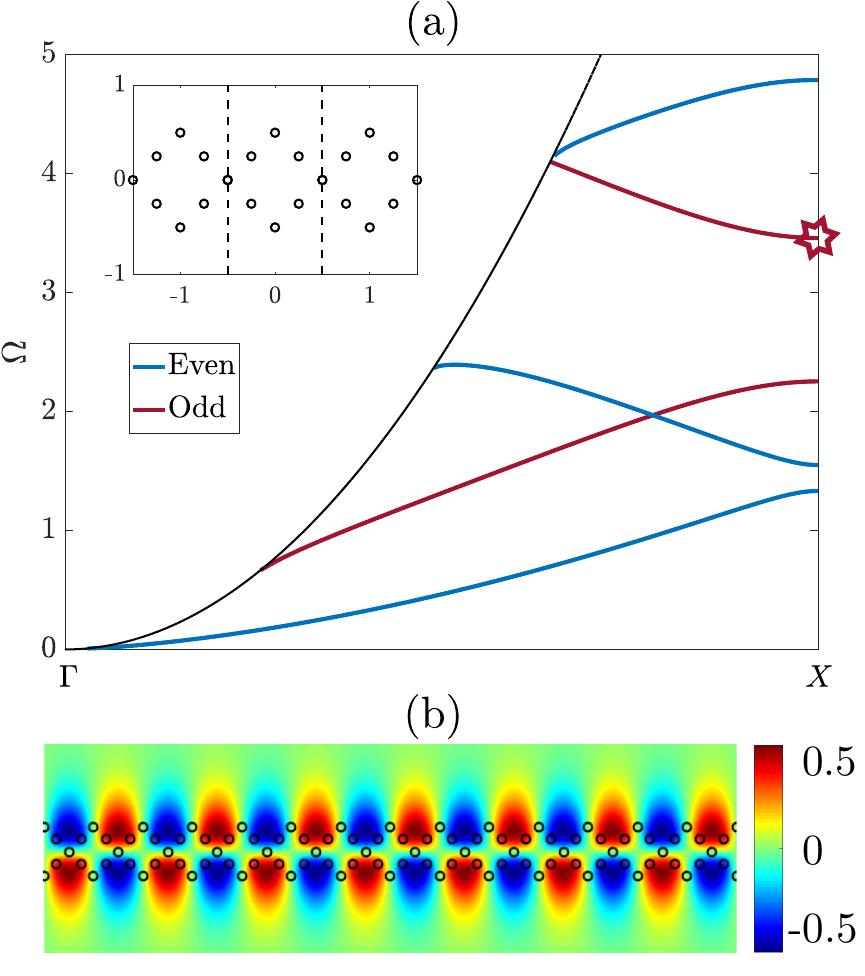}
\caption{\textit{Dispersion curves and eigenmodes:} (a) Dispersion curves for the diamond arrangement of point masses, shown inset, in a strip of width $a = 1$ and masses, $M = 10$, with diamond centroid to vertex distance $r = 0.5$ (coincident masses superimpose). 
Curves corresponding to even, or odd, eigenmodes relative to the line array are coloured blue and red respectively. (b) Real$(w)$ for $\Omega = 3.45$ (starred in (a)). 
}
\label{fig:DispWF}
\end{figure}
An essential element in designing graded arrays, varying adiabatically, is to accurately determine dispersion diagrams for ungraded, perfectly periodic, arrays; we use a  Fourier-Hermite spectral method designed to rapidly extract only the array-guided waves \cite{Chaplain2019}. The advantages of this spectral scheme allows us to avoid fully numerical approaches such as finite elements and perfectly matched layers \cite{Vial2014}, and it enables many geometrical parameters, and their corresponding dispersion curves, to be analysed to find optimal graded composite structures to produce rainbow trapping, reversal of the array guided wave and the hybridisation into the bulk.

Throughout this Letter we take a line array constructed from strips of width $a$ each containing a single shape, and use point masses arranged around a diamond, see Fig. \ref{fig:schematic}. The point masses are placed at its vertices and mid-way between the vertices; the diamond is characterised by a parameter $r$ the distance from the centroid of the diamond to the vertex. In the graded regions we use a linear grading with the diamond parameter $r$ varying by a grading index $\Delta r$   from strip-to-strip. 

 Mode shapes are quickly identified alongside the dispersion relations shown in  Fig.~\ref{fig:DispWF}. The modes naturally fall into those having either odd or even symmetry with respect to the line array; from the symmetry the even and odd modes can be decoupled provided the excitation has the same symmetry. We focus on the second odd mode to demonstrate the effect, and note that other mode branches can be used as detailed in Supp. Mat., there is no overlap between this second odd mode and other modes and this lack of coupling can be utilised for more  conventional mode conversion in structured KL plates \cite{Chaplain2019}.

The diamond structure is then graded by altering the radius of the circle on which the vertices of the diamond lie, a schematic of which is shown in Fig.~\ref{fig:schematic}. The grading profile is determined from dispersion curve analysis (Fig.~S1); detailing at which grading parameters such modes are supported. The dispersion curves are shifted in frequency in a direction determined by the grading profile, shown in  Fig.~\ref{fig:interp}(a). The change in the dispersion curves as a function of grading profile can be used to predict where the trapping and subsequent hybridisation can be achieved (supplementary Figs.~S1,~S3).

\begin{figure}[h!]
\centerline{
\includegraphics[width = 0.45\textwidth]{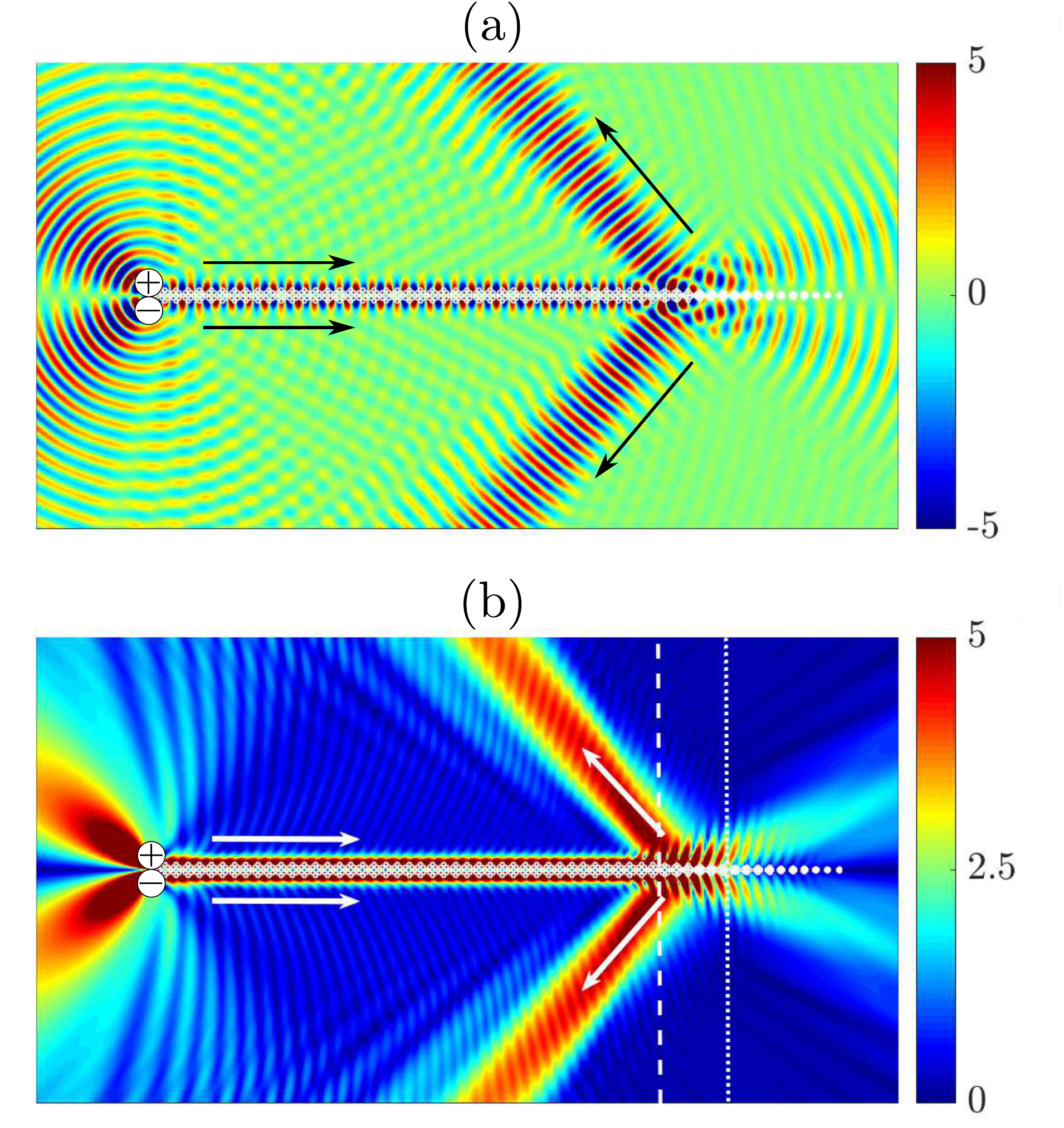}}
\caption{\textit{Reversed Rainbow effect:} (a) Real$(w)$  
 for the array shown in Fig. \ref{fig:schematic} with strip width $a = 2$, mass $M = 10$ forced by a dipole with $\Omega = 1.86$. The array is ungraded, $r=1$, in the central region and graded, with grading index $\Delta r = -0.05$, in the region beyond the dashed white line in (b). (b) Abs$(w)$, with dipole position and polarity shown within white circles.  The dotted white line shows point of initial reflection and rainbow trapping (see also Fig.~S1).}
\label{fig:ReverseRainbow}
\end{figure}

\begin{figure*}
\includegraphics[width = 0.75\textwidth]{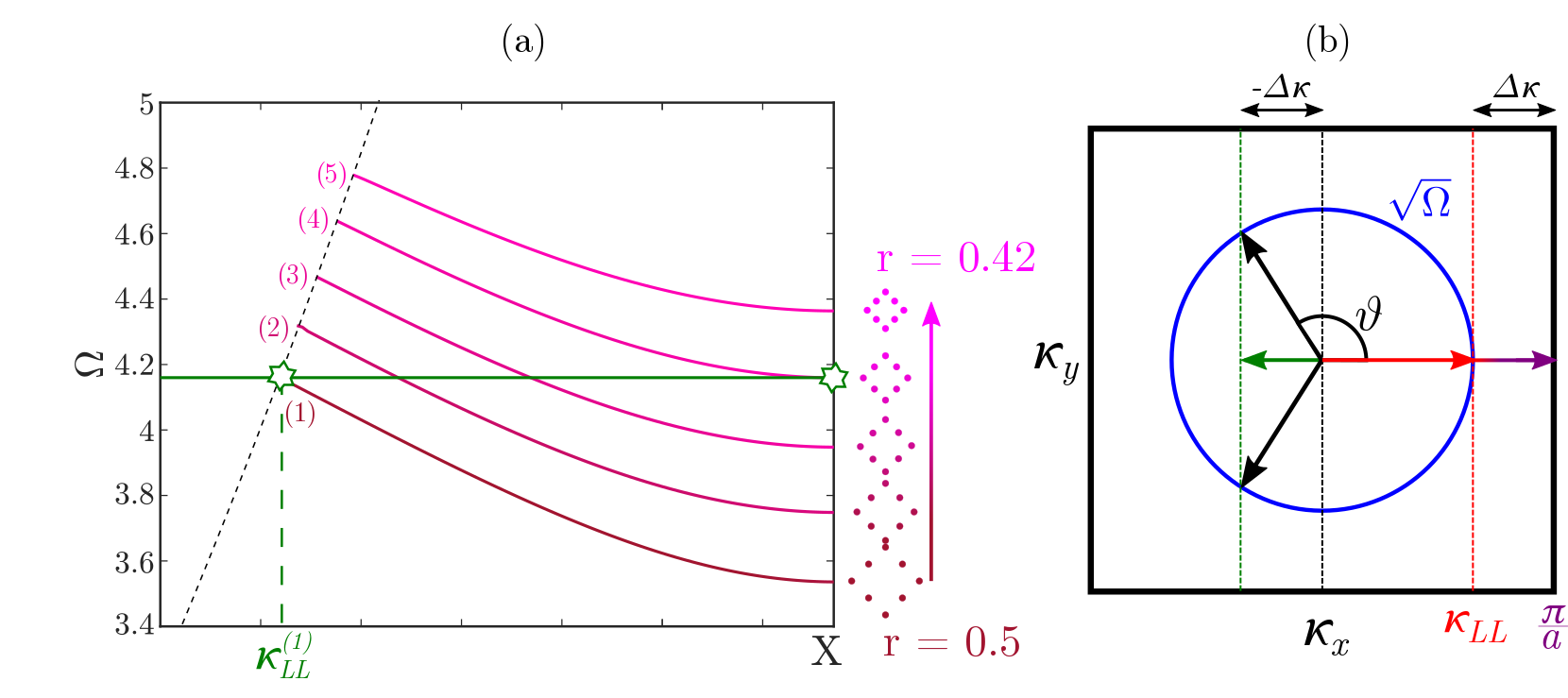}
\caption{(a) \textit{The second odd dispersion curve:} The curve moves upward as the diamond shrinks; the strip has constant width $a = 1$ with masses $M = 10$. The excitation frequency $\Omega = 4.165$ aligns with light line for curve (1)  at $\kappa_{i} = \kappa_{LL}^{(1)}$ and, as $r$ decreases, aligns with curve (4) at $\kappa_{f} = X \equiv \pi/a$. (b) \textit{Wavevectors:} Isofrequency contour of the homogeneous KL plate (blue circle) and wavevector $\kappa_{LL}^{(1)}$ of the AGW (red arrow) in the ungraded region which increases, with grading, to the band-edge until $\kappa_{f} =\pi/a$ (purple arrow) whereupon  
    rainbow trapping  occurs and the wavevector flips, acquiring a phase-shift of $\pi/a$ due to the reflection.  The reverse grading
     decreases the phase by $\Delta\kappa=\kappa_{LL}-\pi/a$ leading to an 
     effective horizontal wavenumber $-\Delta\kappa$ lying within the isofrequency contour of the bulk medium (green arrow). Phase matching gives the hybridisation into the plate, i.e., to waves propagating in the directions shown by black arrows at angle $\theta$ to $x$-axis.}
     \label{fig:interp}
\end{figure*}

The hybridisation of the initially forward propagating array guided mode into backward facing beams, as shown  in Fig.~\ref{fig:ReverseRainbow}, going into the bulk is achieved using ideas that draw upon rainbow trapping \cite{Tsakmakidis2007} together with phase-matching. These trapped and hybridised beams differ to the backward leaky antenna waves of \cite{Cai2015} in their origin and character; they do not arise from losses and exist in the same plane as the structuring. Further to this the spatial position of their generation can be tuned geometrically. 

To demonstrate this effect we begin by reversing the array guided mode and to do so we gently grade the array to induce rainbow trapping. This can be interpreted using   Fig.~\ref{fig:interp}(a) where the dispersion curve for the ungraded array (with initial wave vector $\kappa_{i} \equiv \kappa_{LL}^{(1)}$ of Fig.~\ref{fig:interp}(a)) is the first curve and it moves upward as the grading occurs until, at the fourth curve, the curve touches the band-edge and subsequently the operating frequency is in a band-gap; the array guided mode then reverses direction whilst still being on the array. The mode also acquires a phase shift of $\pi/a$, a momentum `kick', upon reflection and now experiences the opposite grading profile. The upshot of which is that the phase of the reversed graded array mode is $\kappa_i - \pi/a$ as it returns to the ungraded portion; this lies within the isofrequency contour of the bulk medium and, phase matching its horizontal component, allows the energy to hybridize into the backward facing beams that are observed (Fig.~\ref{fig:interp}(b)). The effect can be partially reversed, using reciprocity, with two Gaussian beams fired toward the graded array, at the corresponding angle, to excite a backward propagating AGW, see Fig.~S5.  
 

\begin{figure}
    \centering
    \includegraphics[width = 0.5\textwidth]{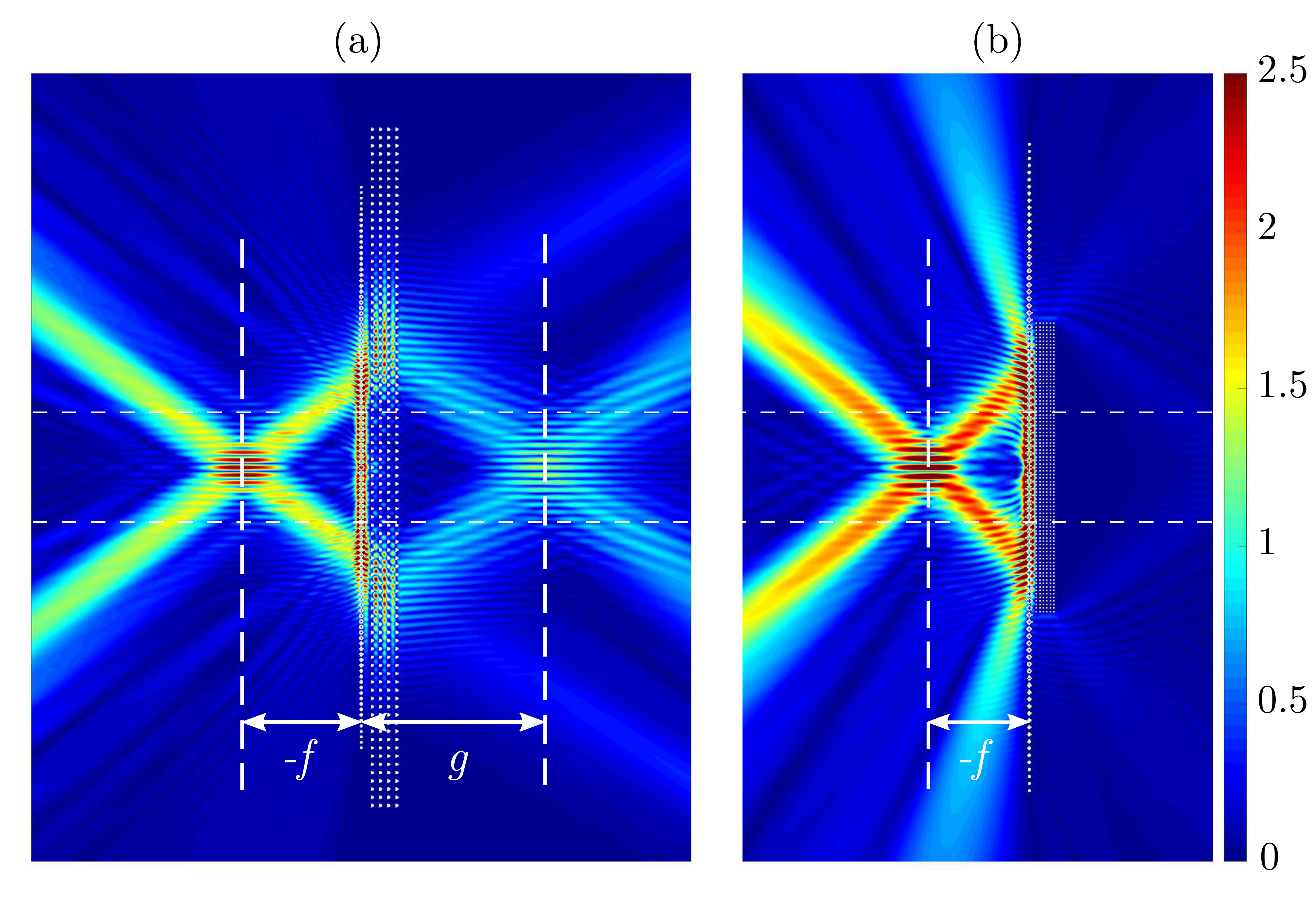}
    \caption{\textit{Generalised Flat Lensing:} (a) Coupling the graded array with a masking phononic crystal such that different focal distances, typical of glenses \cite{Chaplain2016}, arise. (b) One-way lens using a complete frequency mask to the right of the array. Dipole sources ($\Omega = 3.7$) are placed at the array centre with $a = 1$, $M = 10$ with ungraded $r=a/2$ between the horizontal dashed lines and graded outside them  with $\Delta r = -0.01$. Crystal parameters given in Table~S1, Fig.~S11.}
    \label{fig:GenFlatLens}
\end{figure}

Grading the line array symmetrically about the dipole position, as in  Fig.~\ref{fig:FlatLens}, and tuning the frequency, creates two images with equal focal distance $f$ on either side of the array; equivalent to the source being placed at the internal focus of symmetric negatively refracting line. Asymmetric grading about the dipole yields a degree of control over both horizontal and vertical positions of the focal spot, see Fig.~S10. A further degree of control is gained by applying a mask to the array by placing a phononic crystal to one side; the dipole is then imaged to two different focal lengths $-f,g$, see Fig.~\ref{fig:GenFlatLens}(a),  simulating the effect of a source placed off-centre within a negatively refracting lens, and in turn resembling a glens \cite{Chaplain2016}. Engineering the different focal lengths combines classical Snell's law with the properties  of the phononic crystal \cite{chuang2012physics, smith2012negative} (described in detail in the Supp. Mat. and Fig.~S11, Table~S1). The masking effect of the crystal is tuned by altering its components and geometry, for instance by operating in its band-gap, by taking the point masses to be point clamped constraints, a one-way lens can be designed, see Fig. \ref{fig:GenFlatLens}(b).

A common feature of array guided waves, particularly if composed of resonant components, is that they are not propagating but weakly decaying along the array. This decay does not prohibit the application of our concept, i.e. we take the same diamond arrangement and graded structure, but with the point mass-loading replaced by a mass-spring system with resonances. Tuning the resonant frequency pushes the dispersion curves down/up in frequency and flattens them out, effectively narrowing the frequency band over which waves can propagate and encouraging weakly decaying guided waves. The reversal and hybridisation effect still occurs for these decaying guided modes provided the graded region is engineered close to the source. Since we are now dealing with decaying waves along the array the effect is naturally determined by the rate of the decay, which we determine  using High Frequency Homogenisation (HFH) \cite{Craster2010,Chaplain2019} -- an asymptotic scheme which allows the characterisation of decay and envelope wavelengths of AGWs through a multiscale expansion (Figs.~S6,~S7). Coupling the use of resonant elements with the geometrical design paradigm above provides motivation for extension to subwavelength imaging.

In conclusion, we have demonstrated that grading an array, and carefully tuning the grading, not only reverses the direction of the array guided waves, as in rainbow trapping, but can be used to hybridize the reversed array waves into the bulk. This is effectively shrinking negative refraction to occur upon a line array, the resulting beams within the bulk then create focal points; the angle of reversal depends on the excitation frequency and the grading profile and so the focal point position is tuneable; as a result there are a plethora of candidates for the extension to broadband applications.

We chose to illustrate this wave phenomenon via flexural waves propagating within the KL elastic plate model which has many computational advantages; the Green's function is non-singular and this allows for straightforward numerical methods that sidestep finite element or other methods such as multipole or plane-wave expansions. The wave phenomena uncovered are not specific to these elastic flexural plate waves, but are generic to periodic arrays for which grading is feasible.  We anticipate further applications across wave physics, i.e. in optics and electromagnetism where graded metascreens, building on the flat optics work of \cite{Yu2014}, could create classes of flat lenses, with the potential for subwavelength imaging.

The authors thank the UK EPSRC for their support through Programme Grant EP/L024926/1 and a Research Studentship. R.V.C acknowledges the support of the Leverhulme Trust.

\bibliographystyle{apsrev4-1}
%

\newpage
\onecolumngrid

\section{ Flat Lensing by Graded Line Meta--arrays:  Supplementary Material}
\vspace{1cm}

\twocolumngrid

{\it \textbf{Introduction}:}
We provide supporting detail on the effective negative refraction induced on a line array by using rainbow trapping and wave hybridisation.  The design paradigm is introduced via a Fourier--Hermite Galerkin spectral method that efficiently extracts dispersion curves, these curves guide the physical interpretation of the main text. This design is further corroborated by full numerical simulations and by using reciprocity. 

In addition to determining the geometries that exhibit the trapping and hybridisation for propagating array-guided waves we turn our attention to decaying array localised modes that lie in the band-gap of the ungraded array; this is achieved by  recapturing the decaying modes as they propagate along the array. This model is extended to resonant loading, presenting motivation for subwavelength operation. Further to this, more novel optical components, namely the cylindrical mirror, are emulated.

In the main text we place the graded array above a photonic crystal to control the wave behaviour on each side of the array and here we present the array-crystal coupling theory required by examining doubly periodic structures, and their isofrequency contours, and showing their interaction with the array. Further wave manipulation effects that complement those of the main text are also shown.

The line array is constructed from strips of width $a$ each of which contains a single shape, here we illustrate all our examples using point masses arranged around a diamond. The point masses are placed at its vertices and mid-way between the vertices; the diamond is characterised by $r$ the distance from the centroid of the diamond to the vertex. Variation in the properties of this line array occur by altering the spacing between diamonds, i.e. altering the strip widths, or by grading the sizes of the diamonds or by altering the masses. 

\onecolumngrid

\begin{figure}[b!]
	\centering
	\includegraphics[width = 0.95\textwidth]{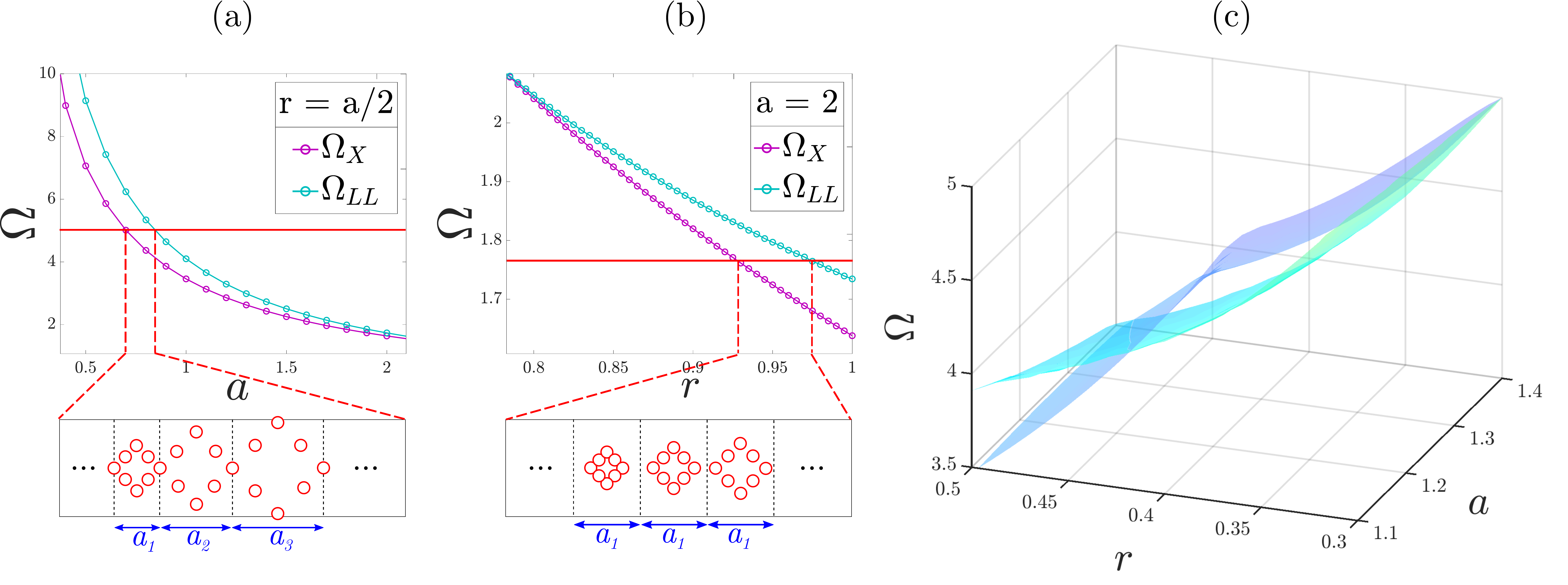}
	\caption*{Figure S1: \textit{Propagation Diagrams:}
		The propagating frequencies for the second odd mode, Fig.~5, the mode used in the main text, are determined by the intercepts $\Omega_{LL}$ and $\Omega_X$, with the ``light-line" (actually a sound-curve for an elastic plate) and band-edge $X$. For constant grading parameter, for 
		frequencies between $\Omega_{X}$ and $\Omega_{LL}$ an array-guided wave is supported. For given frequency, say $\Omega$ the horizontal red line in (a,b), the range in $a$ or $r$ for which the wave is supported lies between the two curves. 
		(a) Effect of varying strip width, $a$, with constant diamond parameter, $r = a/2$,  with a schematic of grading profile below. (b) Effect of varying diamond parameter $r$, with constant strip width $a$ and grading profile below. (c) Surface plot for varying both diamond parameter and strip width; the point masses have $M = 10$. The lines/surface which are purple or blue correspond to $\Omega_{X}$ and  $\Omega_{LL}$ respectively.}
	\label{fig:SurfProp}
\end{figure}
\twocolumngrid
{\it \textbf{Kirchhoff-Love equations and numerical approach}:}
The (non-dimensionalised) Kirchhoff-Love equation governs the flexural wave modes that exist on an infinite elastic plate with point constraints at lattice points, characterised by their vertical displacement $w(\mathbf{x})$. Implicitly assuming a time harmonic response, this equation is shown in the main text (Eq. (1)) for simple mass-loading. Below we show the case for resonant loading by incorporating point mass-spring resonators, each of mass $M$ with spring constant $k$ so that
\begin{align*}
F(\mathbf{x}) = -k\sum\limits_{N}\sum\limits_{p =1}^{P}[w_{N}^{(p)}(\mathbf{x}) - \Tilde{w}_{N}^{(p)}(\mathbf{x})]\delta(\mathbf{x} - \mathbf{x}_{N}^{p}), \tag{S1} \\
-\Omega^2M\Tilde{w}_{N}^{(p)}(\mathbf{x}) = k[w_{N}^{(p)}(\mathbf{x}) - \Tilde{w}_{N}^{(p)}(\mathbf{x})], \tag{S2}
\label{eq:res}
\end{align*}
where we have introduced the unit strip labelled by $N$, which repeats periodically in the $\hat{\mathbf{x}}$ direction. Each strip contains $P$ resonators, whose displacement is given by $\Tilde{w}_{N}^{(p)}$.

The governing equation can be solved to obtain the eigenstates using a variety of spectral methods, including plane wave expansions or the Fourier--Hermite method as used in the main text. Alternatively, a Green's function approach \cite{evans2007penetration,torrent13a} can be used and the total wavefield, for $Q$ scatterers, can be obtained by evaluating
\begin{equation}
w({\bf x}) = w_s({\bf x}) + \sum_{q=1}^{Q} F_q g\left(\Omega, |{\bf x}-{\bf x}_q|\right), \tag{S3}
\end{equation} 
where $w_s({\bf x})$ is the input field which will, for example, encode point forcings, as used in the main text. 

Using the well-known Green's function \cite{evans2007penetration},
\(
g\left(\Omega, \rho\right) = ({i}/{8\Omega^2})
\left[H_0(\Omega \rho) - H_0(i\Omega \rho)\right]
\),
the unknown reaction terms $F_q$  come from the linear system
\begin{equation}
F_m = M_m\Omega^2\left[w_s({\bf x}_m) + \sum_{q=1}^{Q}
F_q g\left(\Omega, |{\bf x}_m-{\bf x}_q|\right) \right]. \tag{S4} 
\end{equation} 
Due to the attractive features of the Green's function for this setting, these calculations are straightforward to implement and allow full scattering solutions to be computed for corroboration to the results achieved from the spectral methods employed.

\begin{figure}[h]
	\centering
	\includegraphics[width = 0.4\textwidth]{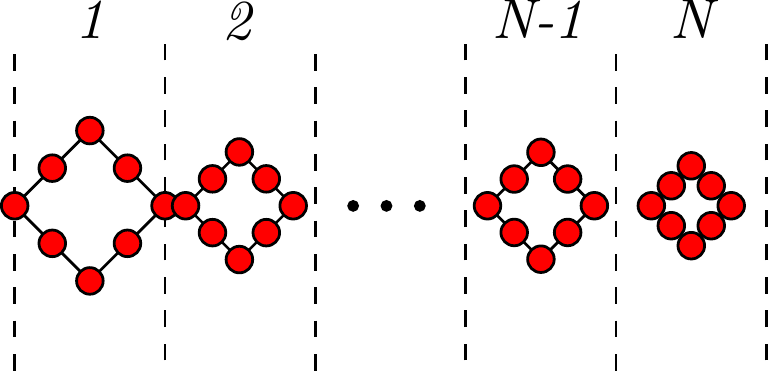}
	\caption*{Figure S2: \textit{Example of graded region and cell numbering.}}
	\label{fig:GradeArg}
\end{figure}

\begin{figure}
	\begin{tabular}{p{0.05\textwidth} p{0.4\textwidth}}
		\includegraphics[scale = 0.22125]{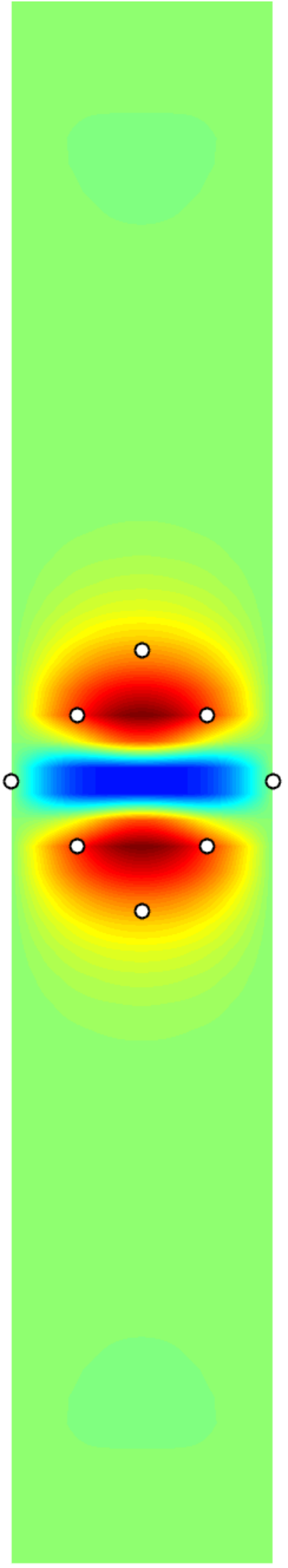} &
		\includegraphics[width = 1.1\linewidth]{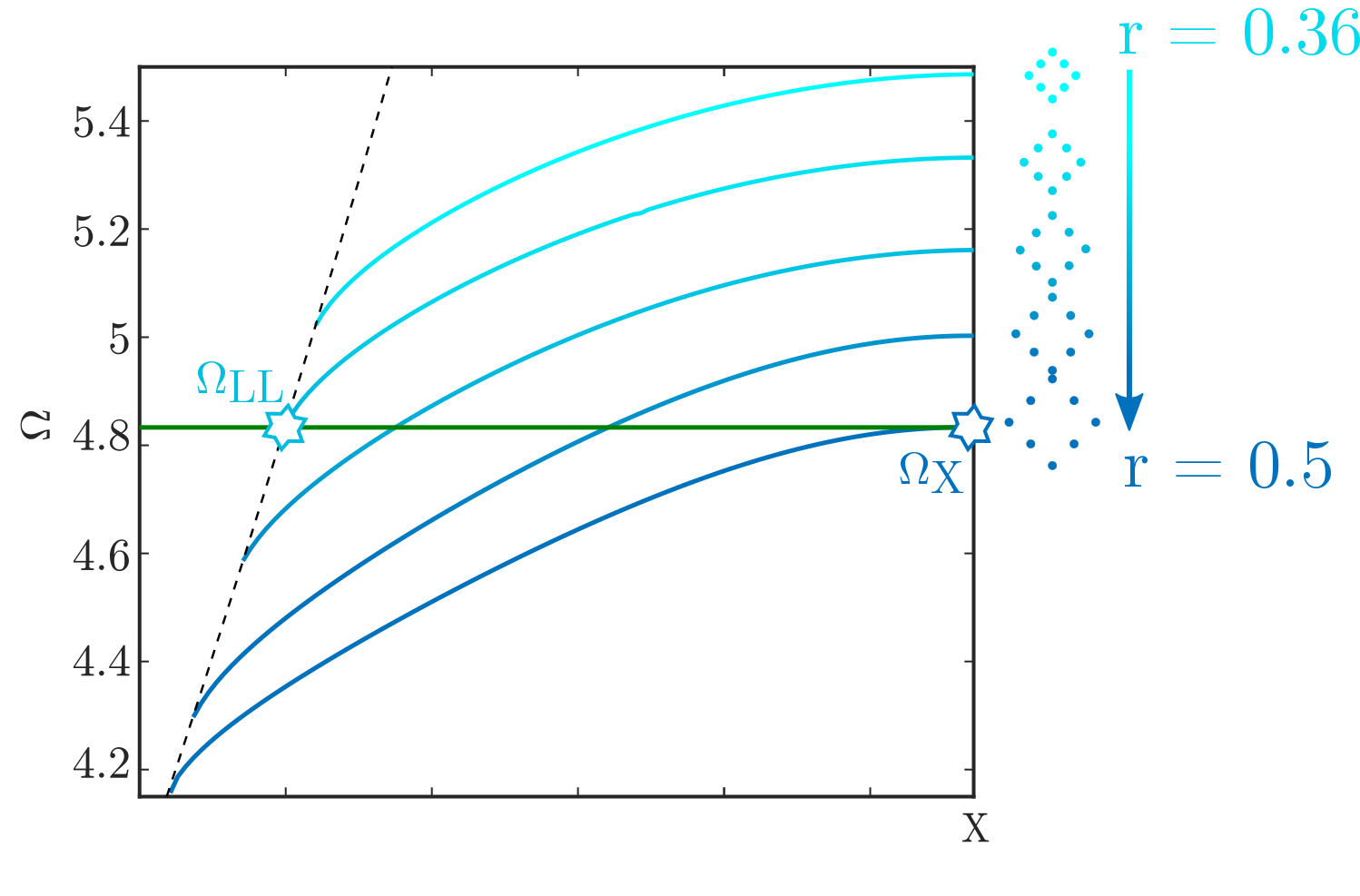}
	\end{tabular}
	\caption*{Figure S3: \textit{Dispersion curves for the third even mode  for differing
			$r$:} Each curve represents the third even mode 
		for increasing $r$ for strip width of
		$a = 1$ ($M = 10$). We deliberately choose an even mode, rather than the odd mode of the main text in order to 
		contrast with Fig.~5. Here due to the positive slope of the third even curve, to achieve the rainbow trapping the grading parameter, $r$, has to \textit{increase} in order for the excitation frequency ($\Omega = 4.833$) to initially align with the light line (for $r = 0.36$) and then with the band edge at $r = 0.5$. A schematic of the grading is shown, with graded colour scheme, alongside the curves along with the real part of the wavefield at $\Omega=4.833$ showing the mode is even.}
	\label{fig:PositiveSlope}
\end{figure}

{\it \textbf{Design strategy}:}
To design a graded array to support array guided waves that are first rainbow trapped, reverse direction, and then hybridised into a bulk mode, we exploit the technique developed in \cite{Chaplain2019} for extracting the modes that propagate along an ungraded array and exponentially decay perpendicular to the array. A Fourier-Hermite spectral method exploits the periodicity along the array, through the Fourier expansion, and automatically builds in decay through the Hermite expansion, to arrive at a spectrally accurate method to extract only the array guided waves. This allows us to extract dispersion curves for the ungraded array and to rapidly characterise the effect of varying either strip width or diamond size, see Fig.~S1. 

Given a choice of  grading parameter, this could be the centroid to vertex inclusion distance, $r$ (as in main text), the width of the unit cell, $a$, or the mass/resonant parameters of the point loads or the shape around which the inclusions are placed, the spectral method is employed to  generate the dispersion curves for each variation in grading. From the dispersion curves, such as Fig.~3, individual modes are identified, and the frequencies where they meet the light line, $\Omega_{LL}$,  and band edge, $\Omega_X$, are determined, see Fig.~S1 that shows the frequency range over which the second odd mode exists. 

`Propagation surfaces' can be realised by the multi-grading of, for example, strip width and geometry within the strip, as shown in Fig.~S1(c). For the frequencies where the dispersion curves cut the light-line (blue) and the band edge (purple) in Fig.~S1 a horizontal slice at a specific frequency gives the range of grading parameters, for this mode, for which an array-guided wave propagates.
Tuning the array properties spatially to ensure that the grading incorporates these values enables us to create rainbow trapping and reverse the direction of the wave, and subsequently engineer the mode to propagate into the plate along a direction of our choosing.

For clarity we focus on grading the diamond parameter $r$, as in the main text, and show how increasing or decreasing $r$ by the grading index, $\Delta r$, in subsequent cells along the array is used in the design process. The mode used in the main text, focused on purely spatially odd waves about $y$, i.e. relative to the line array, see Fig.~3(b); this is not a necessity for the array guide wave reversal, and hybridisation, and even modes can also be used.

\begin{figure}
	\centering
	\includegraphics[width=0.45\textwidth]{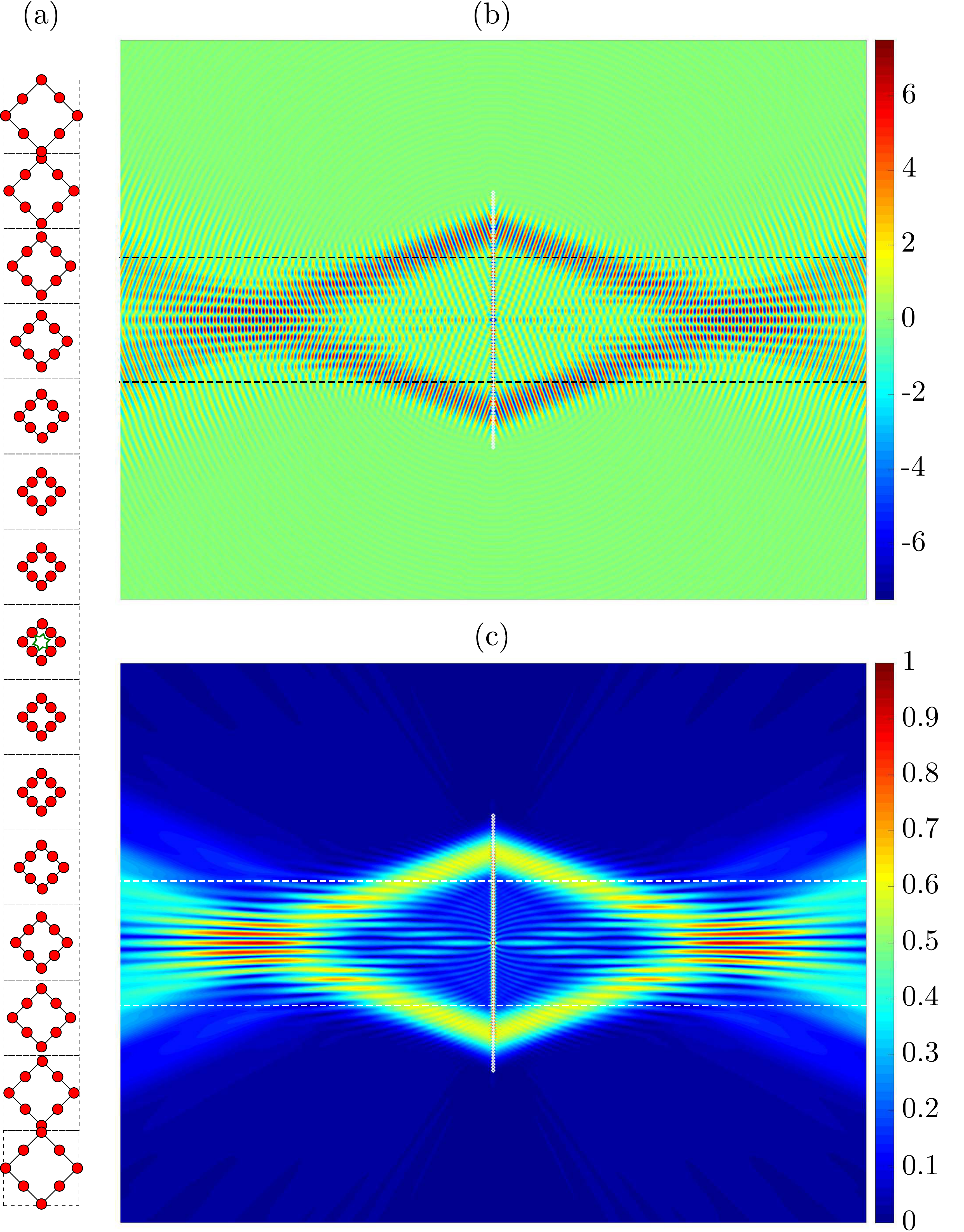}
	\caption*{Figure S4: \textit{Lens effect for third even dispersion curve:} (a) Schematic of array grading using, i.e. increasing $r$, with monopole source position shown by green star in centre of lens; excitation frequency is $\Omega=4.833$. (b-c) Real/Absolute part of wavefield demonstrating  flat lensing for the even modes in Fig.~S3. Grading regions are beyond dashed horizontal lines, with $\Delta r = +0.01$. Arbitrary units are used throughout for the colourbars, as in main text.}
	\label{fig:EvenLens}
\end{figure}

Consider a graded region, consisting of $N$ unit strips, labelled $1, \hdots, N$ as in Fig.~S2. To achieve rainbow trapping, reversal and hybridisation, the excitation frequency must be near the light line for the initial grading parameter, i.e.  $\Omega \sim \Omega_{LL}$, where $\Omega_{LL}$ is the frequency of the first curve in the graded region to meet the light line, see Fig.~S3. By gradually grading the array by changing $r$ then, for some $N$, at the $N^{th}$ cell  this curve meets the band edge at the excitation frequency, i.e. $\Omega \sim \Omega_{X}$, as shown in Fig.~S3.  We then arrive at the condition relating these two curves, with different $r$, as 
\begin{equation}
\Omega_{LL} \sim \Omega_{X}, \tag{S5}
\label{eq:condition1}
\end{equation}
and thereby implicitly the grading required. It also informs the direction of grading (e.g increase or decrease of $r$ along the array), based on the sign of $\Omega_{LL} - \Omega_{X}$ for each band. In the main text, and Fig.~S1, we consider an odd mode with $\Omega_{LL} > \Omega_{X}$ and in that case a decrease in $r$ ensures (\ref{eq:condition1}). For the third even curve,  Fig.~S3, $\Omega_{LL} < \Omega_{X}$ and so an increasing grading profile  is instead required;  exciting this even mode requires an even (say, monopolar) source, see Fig.~S4.  These grading choices are conveniently summarised in the propagation plots of Fig.~S1; regions where the sign of $\Omega_{LL} - \Omega_{X}$ reverse are easily identified, and the grading is then designed accordingly. It is worthwhile  noting that if condition (S5) is a strict equality we have the strongest coupling; all the energy is reflected and converted. If there is not an exact match between $\Omega_{LL}$ and $\Omega_{X}$, then some energy leaks forwards off the array, as seen in Fig.~4.

\begin{figure}
	\includegraphics[width = 0.4\textwidth]{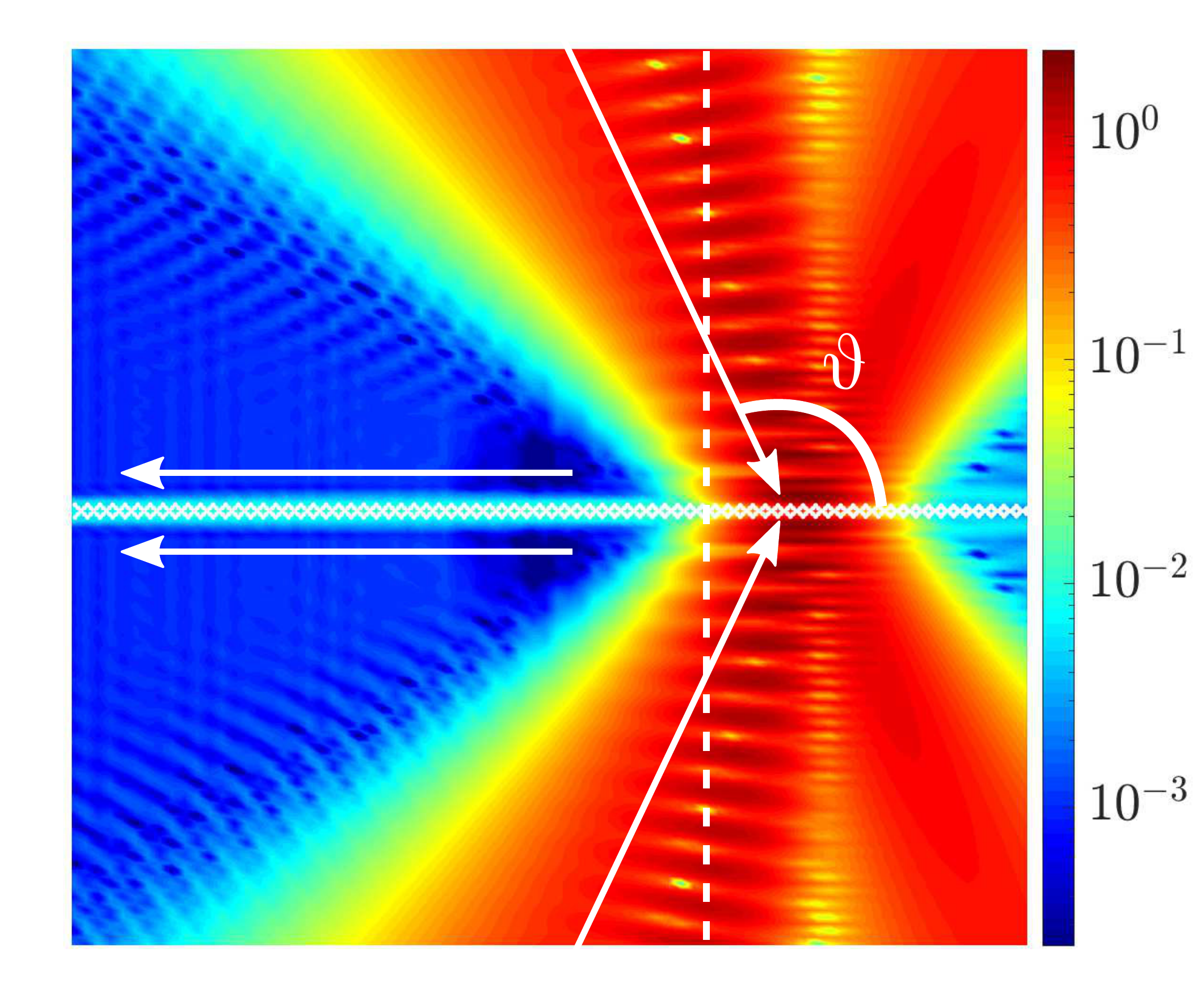}
	\caption*{Figure S5: \textit{Approximate Reciprocity:} Logarithmic scaling of Abs($w$) of Incident and scattered field from two incident Gaussian beams, centred on graded region, at angles $\theta = \pm 114^{\circ}$ with $\Omega = 4.038$.  Strip width $a = 1$ and $M = 10$. Constant region left of dashed line with $r=a/2$, whilst beyond the array is graded with $\Delta r = -0.05$. }
	\label{fig:reciprocity}
\end{figure}

{\it \textbf{Reciprocity}:}
\label{approx}
As a cross-validation of the wave reversal, and subsequent hybridisation into a directed beam on the elastic plate, we approximate reciprocity (we cannot precisely reverse the outgoing beams and instead approximate them as Gaussian beams) and here focus two Gaussian beams on the graded array. Upon excitation, at the predicted angle of incidence, a guided array mode is excited with opposite $\kappa_{x}$ component to the incoming beams, see Fig.~S5. The Gaussian beams are implemented as in \cite{maradudin1990enhanced,Schultz94} with beam amplitude $B(x,y)$ 
\begin{equation}
B(x,y) = \int\limits_{-\pi}^{\pi}\frac{W}{\sqrt{\pi}}e^{-W(\theta - \theta_{0})^2}e^{i\kappa(x\cos\theta + y\sin\theta)}d\theta, \tag{S6}
\end{equation}
where $\theta_{0}$ is the incident angle with respect to the $x$-axis. Motivated by the analysis of Gaussian beams incident on phononic clusters in \cite{smith2012negative}, we choose the beam width $W\simeq 2\lambda$ (where $\lambda$ is the wavelength) so the beam is tightly focused on the graded region, achieving an approximate reciprocal effect.

{\it \textbf{Exploiting decaying array guided states}:}
Thus far we have considered array guided propagating waves, but one can easily operate in the band-gap for the propagating waves and have a mode localised along the array, but decaying, possibly slowly, along it as shown in Fig.~S6(a). 
For frequencies just outwith the second odd dispersion curve, that is in  the band gap, we can extract the decay using high frequency homogenisation (HFH) \cite{Craster2010,antonakakis12a}; a multi-scale technique that recasts the governing equation into a homogenised partial differential equation. 

\begin{figure}
	\centering
	\includegraphics[width = 0.4\textwidth]{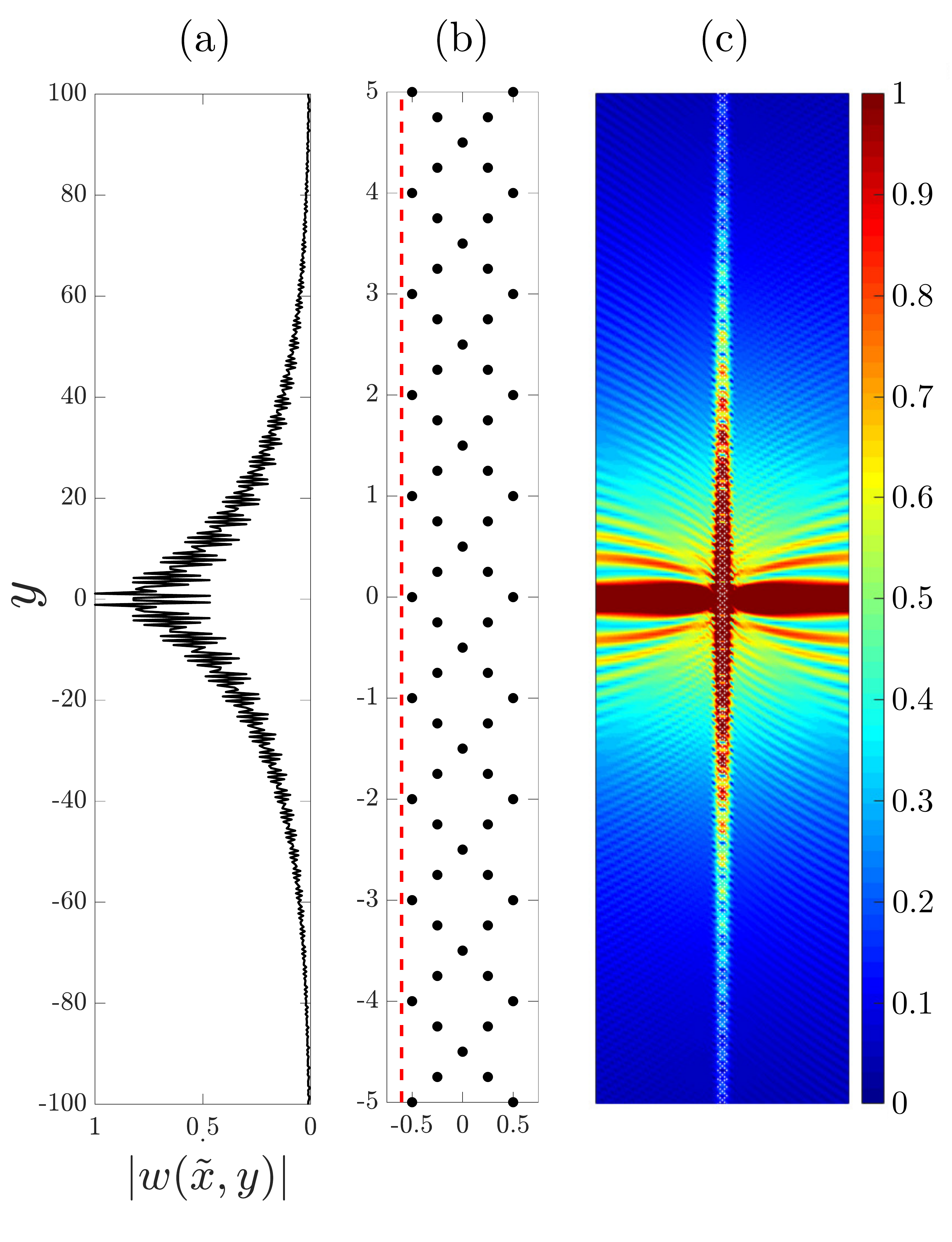}
	\caption*{Figure S6: \textit{Decaying array localised mode for excitation in band-gap:} (a) cross-section of normalised absolute value of field for $\Omega = 4.5$, for an array with constant diamond parameter $r$.s The line along which this cross-section is taken i marked with red dashed line in (b). (c) Shows entire field with prominent beaming from dipole location. Strip parameters are $a = 1$, $M = 10$, $r = 0.5$. }
	\label{fig:DecayArg}
\end{figure}
Focussing on the second odd dispersion curve, as in Fig.~3, an exploded view of which is shown in Fig.~S7, HFH is used, just keeping the first order expansion. 
Strictly speaking, we may not be in a perfect band-gap i.e. there may be a propagating mode of opposite polarity, in this case, even into which the odd field cannot couple.
Fig.~S6(c) shows the absolute value of the field, with prominent beaming perpendicular to the array since in the band-gap much of the energy is directed into exciting a directed propagating plate mode. A decaying mode is excited along the array, with decay length predicted as outlined through the HFH of the semi-infinite dispersion curves (Fig.~S7) and \cite{Chaplain2019}. 

If graded regions are engineered to lie within this decay length then the decaying array field is recaptured by the grading and redirected back towards the dipole, but due to the grading is again rerouted into the bulk.  A triple beaming effect is  observed, see Fig.~S10(b) where the graded array is coupled with a masking phononic crystal, such that there is no excitation to the right of the array.
\begin{figure}[hb!]
	\centering
	\includegraphics[width = 0.4\textwidth]{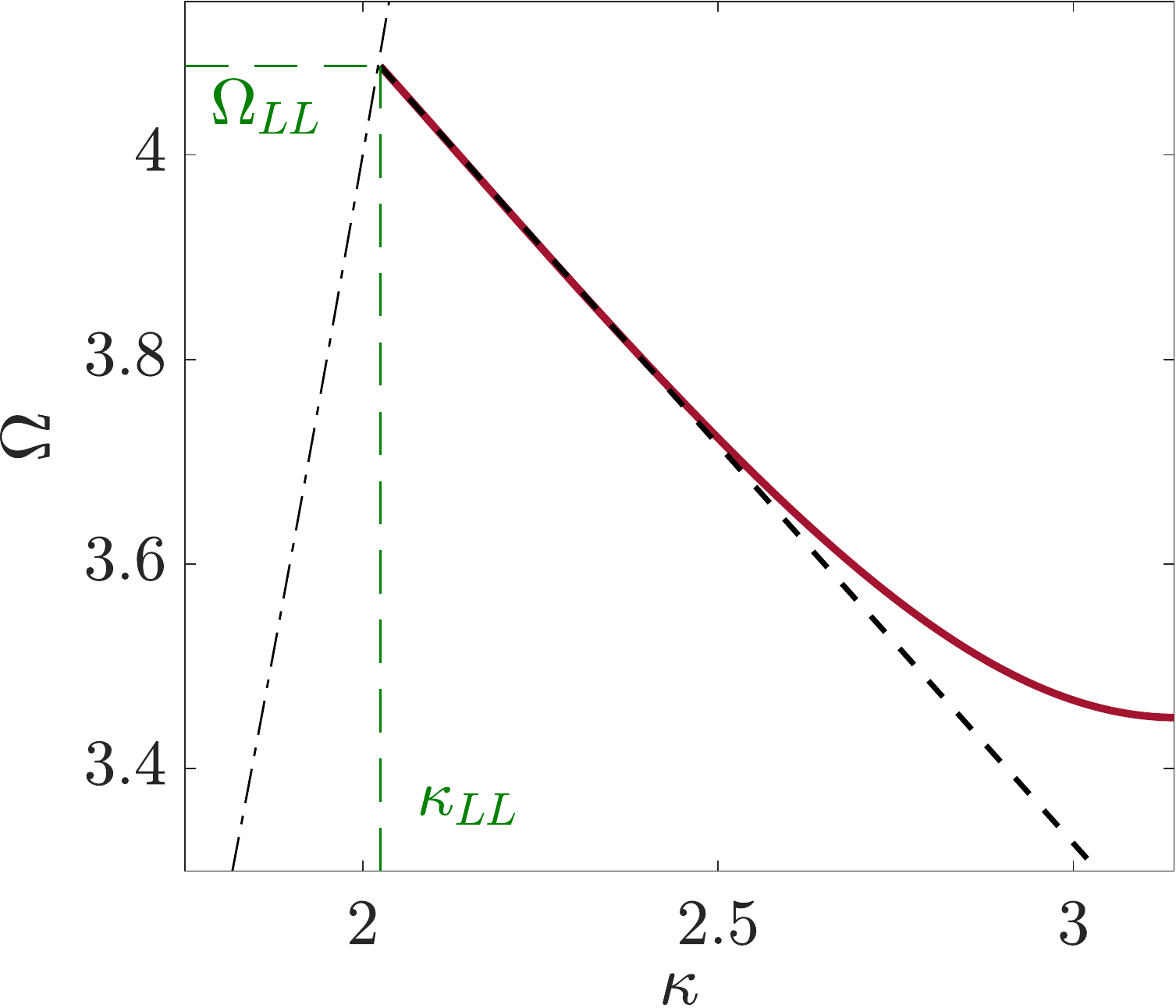}
	\caption*{Figure S7: \textit{High frequency homogenisation (HFH):} Zoom-in of the second odd  dispersion curve from Fig.~3. Dashed black line shows the result from expansion around $(\kappa_{LL},\Omega_{LL})$ using HFH. The slope of the dispersion curve as it approaches the light--line (dash-dotted black line) is linear so only the first order analysis is required.
	}
	\label{fig:HFH}
\end{figure} 

When considering resonant loading, with forcings prescribed by (S1), (S2), these decaying modes are commonplace due to the flattening effect on the bands the addition of resonators often has. We have analysed such mass-spring systems and, similarly to the above analysis, by tuning the grading profile to lie withing the extent of the decay, the frequency range of this effect can be extended. A caveat when operating outside a band is that a strong beam occurs perpendicular on either side of the array, emanating from the dipole source position as a propagating plate mode is also excited. Typical focussing that we achieve is shown in Fig.~S8, where a symmetric grading is used and, due to the resonant elements, weakly subwavelength imaging is achieved. This provides motivation towards further subwavelength imaging flat lenses.

\begin{figure}[t]
	\centering
	\includegraphics[width = 0.425\textwidth]{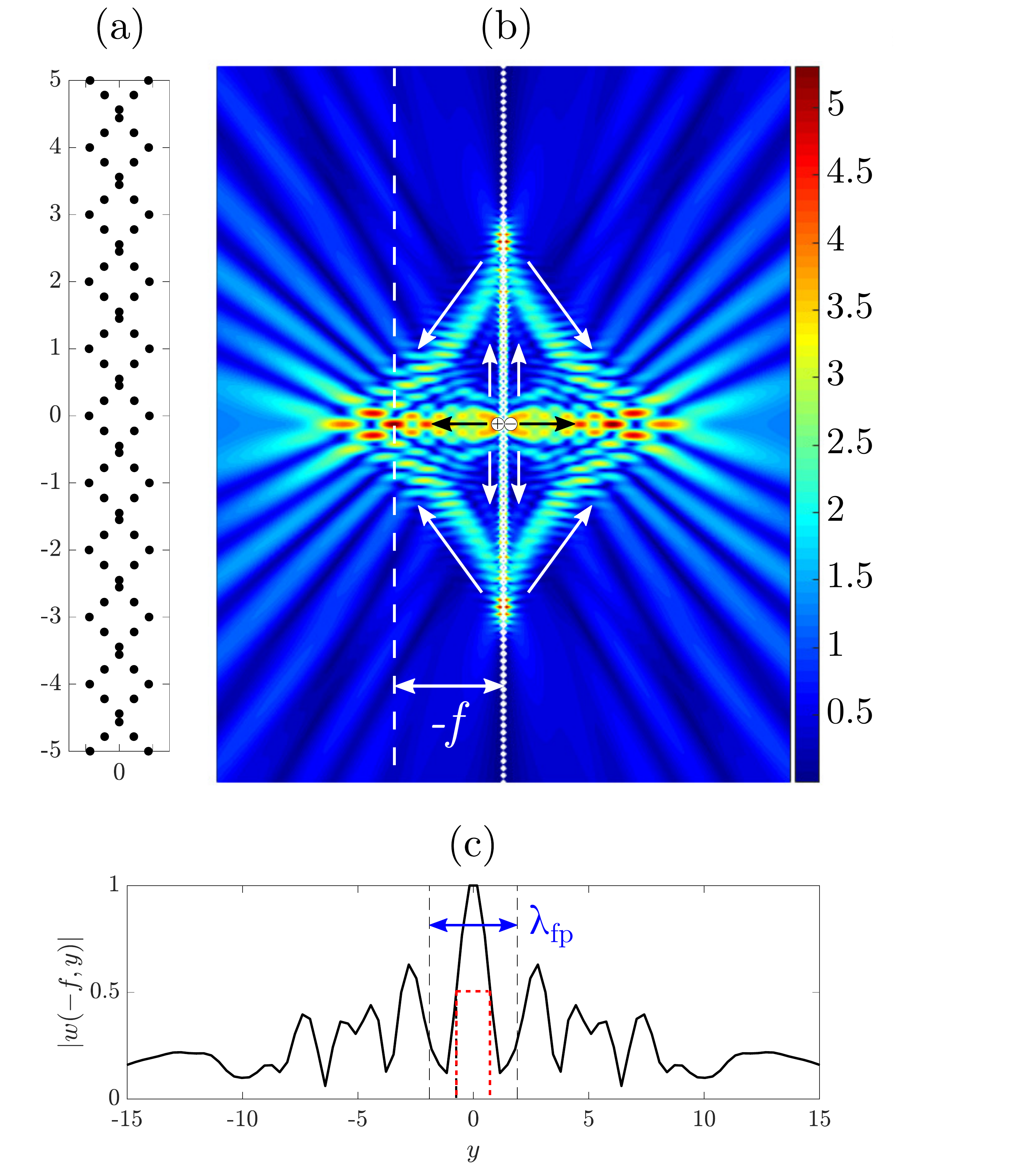}
	\caption*{Figure S8: \textit{Subwavelength Imaging:} (a) Detail of graded array, with $a = 1$ where at the centre $r=0.45$, and grading is immediate with $\Delta r = -0.005$. The array elements are point mass-spring resonators with mass $M = 10$ and spring constant $k = 100$. (b) Abs$(w)$ with dipole excitation (position and polarity denoted by white circles) at frequency $\Omega = 2.71$ (within band gap of AGW). Black arrows show the propagating plate mode excited while white arrows show the excitation of decaying AGW along the array.
		(b) Normalised Abs$(w)$ along $x = -f$ (white dashed line) showing FWHM of focal spot is approximately half of the free plate wavelength, $\lambda_{fp} = 2\pi/\kappa$. } 
	\label{fig:ResRec}
\end{figure}

{\it \textbf{Novel ray optical effects}:}
Further applications, motivated by conventional ray optics, emulate more novel optical components all be it within the context of the phononic elastic wave system we consider, for instance, coupling parallel graded arrays operating as we design allows us to mimic a cylindrical mirror. Fig.~S9 shows the excitation of two identical parallel graded arrays, consisting of an ungraded region close to the sources and a graded region, with differing orientations of point dipole sources indicated by the strengths $\{+,-\}$ or $\{-,+\}$.  The interaction of the reversed direction beams, initiated through  array guide mode reversal followed by hybridisation, means that the focal spot position is frequency dependent, as the angle of beaming from the array is.  These parallel arrays can be interpreted to behave as light rays incident upon cylindrical mirrors, and the excitation frequency mimics the radius of curvature of the analogous ray optical system.

\begin{figure}[h]
	\includegraphics[width = 0.45\textwidth]{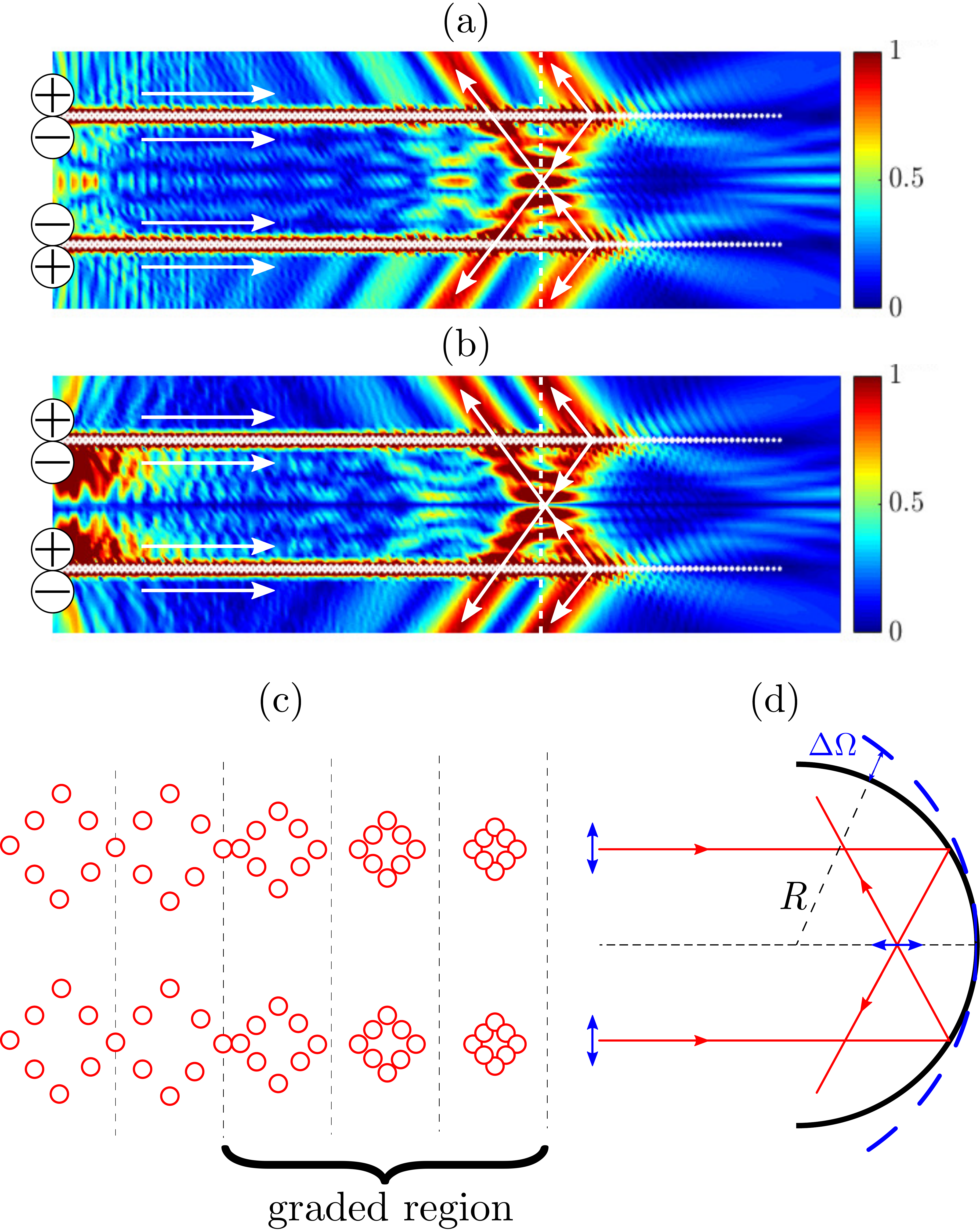}
	\caption*{Figure S9: \textit{Cylindrical Mirror:} (a-b) focusing of reversed and hybridised AGW from two parallel arrays with dipole excitation orientations ($\{+,-\}, \{-,+\}$) and ($\{+,-\}, \{+,-\}$) as marked at array edges within white circles. Strip parameter $a = 1$ with $M = 10$ and excitation frequency $\Omega = 3.7$. The ungraded region has $r=a/2$ and the graded region is denoted with vertical dashed lines such that $\Delta r = -0.01$. (c) Zoom-in near grading region with constant $r$ region shown. By analogy to interior reversed beams, the parallel graded arrays are analogous to the parallel rays in (d) for cylindrical mirror focusing. Altering the array separation and grading, or changing excitation frequency, is analogous to changing the radius of curvature (R) of the mirror and thereby also to tuning the focal spot position.}
	\label{fig:spherical_mirrors}
\end{figure}

{\it \textbf{Phononic crystal coupling and glenses}:}
Inspired by research on placing periodic arrays on photonic/phononic crystals to create negative refractive effects \cite{maradudin2011structured}, we demonstrate the ability to tune the horizontal focal lengths on either side of the line array allowing, in effect, for a generalised lens (glens) to be designed. We couple a doubly periodic phononic crystal, whose unit cell is shown in Fig.~S11(a), to one side of the array; for the angle of incidence of the beam from the line array on the crystal this crystal exhibits negative refraction; this is not essential for the glens effect, however it helps exaggerate the difference in focal lengths. A similar analysis to  \cite{smith2012negative} is used to determine the incident angle at which negative refraction occurs. The isofrequency contours of the free plate and phononic crystal structure are shown in Fig.~S11(b). Utilising the conservation of the transverse component of the wavevector at the interface between the array and the crystal, the direction of the refracted beam is determined. As discussed in the main text, this effect is not achieved due to the local periodicity of the array grading, but by the tuning of  beam incident from the array upon the crystal. 

To then convert the system into a one-way masking lens, a simpler phononic crystal is used, that of a square array whose mass value is tuned as to emulate a self-pinned array, thus preventing any propagation to one side of the array. A summary of the parameters used is shown in Table~S1, with schematics of the array and crystal arrangement shown in Fig.~S11(c,d).
\begin{figure*}
	\centering
	\includegraphics[width = 0.75\textwidth]{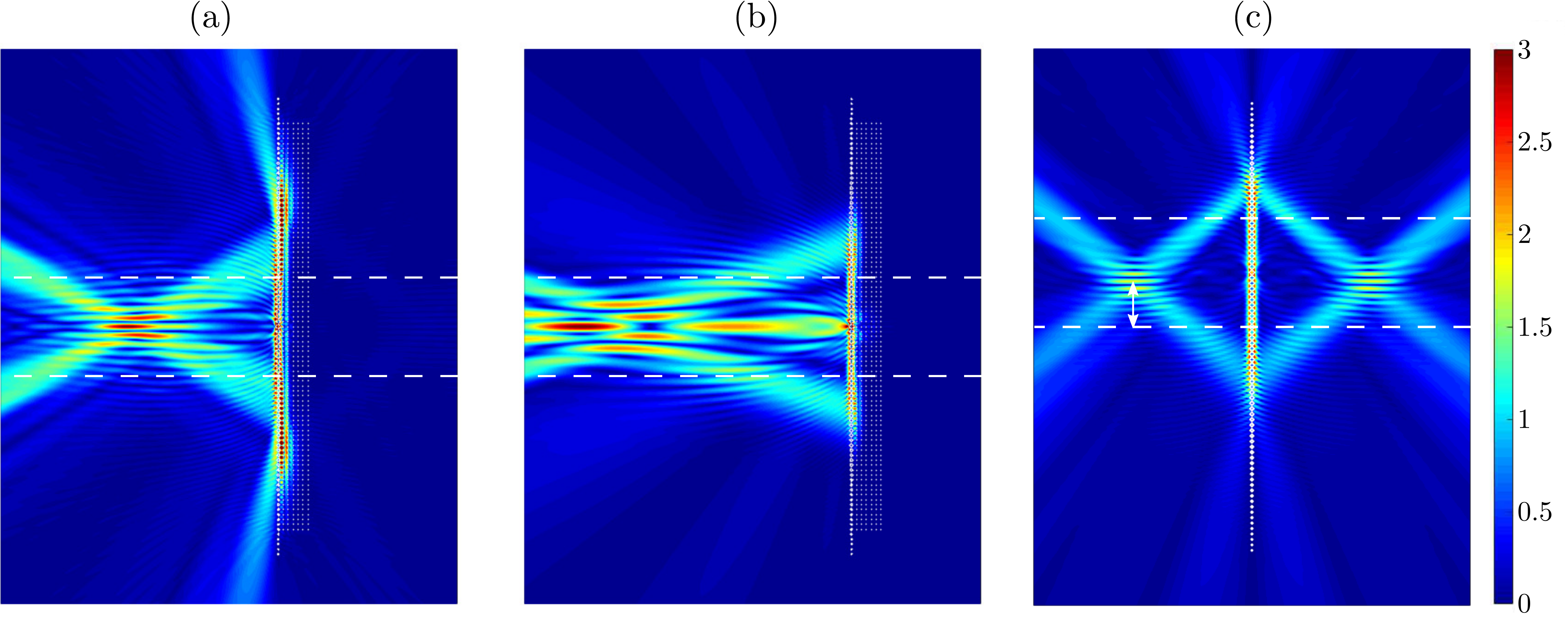}
	\caption*{Figure S10 \textit{Generalisations:} (a,b) have the array backed by a phononic crystal acting as a mask and excitation is by a dipole at frequencies in the band-gap: (a) $\Omega = 4.1$ (b) $\Omega = 4.5$. Both cases have the same array  parameters as in Fig.~S6. In graded regions beyond dashed white lines the grading index $\Delta r = -0.01$. The change in frequency shows how confining the energy by moving further into the band gap alters the directional  beaming perpendicular to the array additionally with a shift in horizontal position of the focus clearly visible. (c) \textit{Off-axis imaging:} Asymmetrically grading the upper and lower halves of the array produces off-axis images, shown by the displaced focal spot (white arrow). Array  parameters are as in Fig.~S5, with $\Omega = 3.7$. Lower graded region corresponds with array midpoint. The graded regions past the dashed white lines have $\Delta r = -0.02$ and $\Delta r = -0.01$ for the upper/lower regions respectively.}
	\label{fig:Decay2freq}
\end{figure*}

\begin{table}[htp]
	\begin{center}
		\begin{tabular}{|| *3{>{\centering\arraybackslash}m{2cm}| } @{}m{0pt}@{}}
			\hline
			& glens & one-way lens \\ [0.5ex] 
			\hline\hline
			Basis vectors & $\mathbf{a}_{1} = [1.5,0]$ $\mathbf{a}_{2} = [0,1.5]$ & $\mathbf{a}_{1} = [0.5,0]$ $\mathbf{a}_{2} = [0,0.5]$  \\ 
			\hline
			Inclusion geometry & \centered{Triangular} &  \centered{Point} \\
			\hline
			Distance of vertices from centroid & 0.2 & 0 \\
			\hline
			Inclusion mass & 15 & 50 \\
			\hline
			Crystal dimensions & $4\mathbf{a}_{1}\times 84\mathbf{a}_{2}$ & $6\mathbf{a}_{1}\times 84\mathbf{a}_{2}$ \\
			\hline
		\end{tabular}
		\caption*{Table S1: \textit{Phononic crystal parameters for examples shown in main text in Fig.~6}}
		\label{tab:PlC}
	\end{center}
\end{table}

\begin{figure}[H]
	\centering
	\includegraphics[width = 0.45\textwidth]{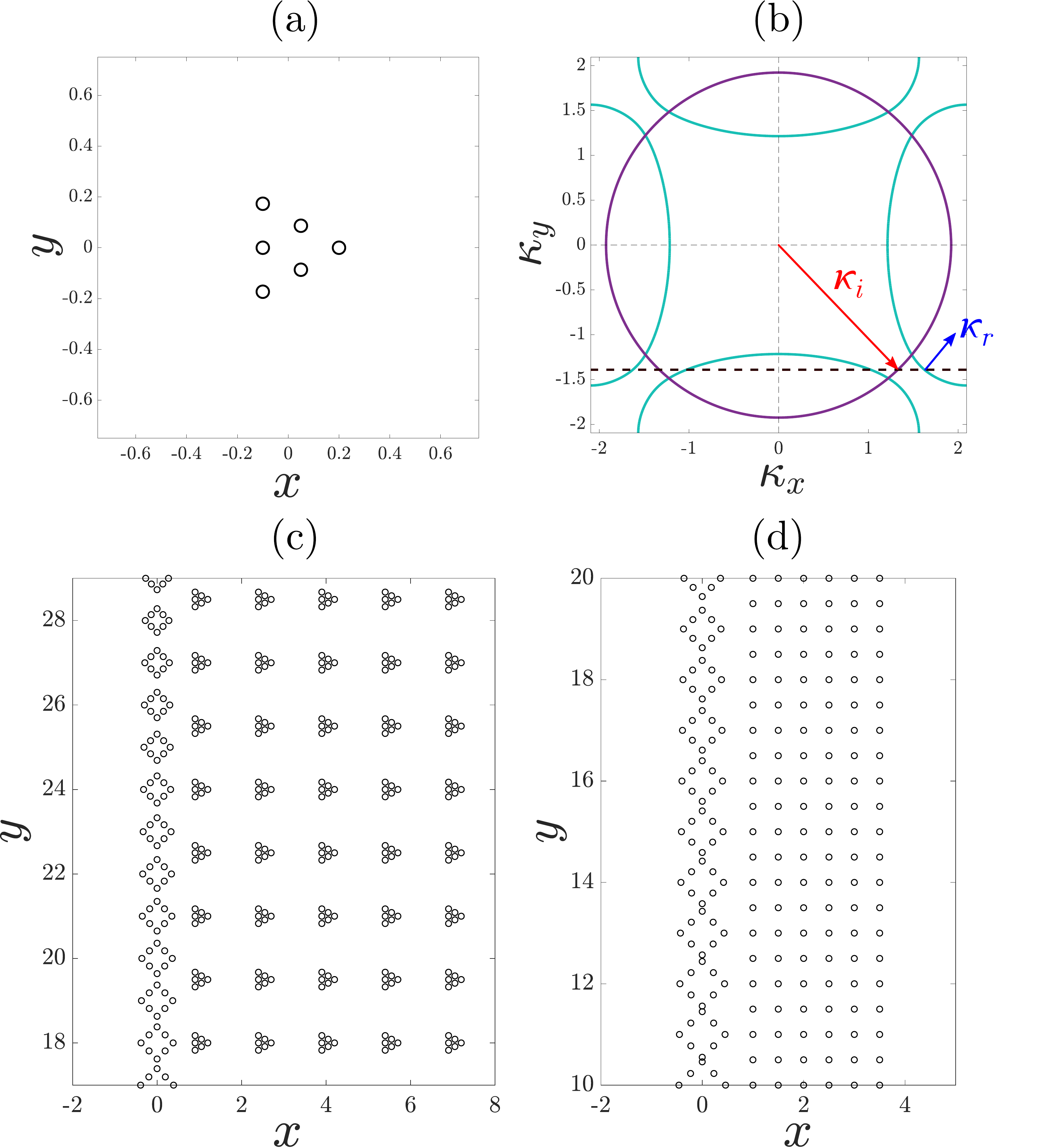}
	\caption*{Figure S11: \textit{Phononic crystals} (a) Unit cell for phononic crystal. (b) Isofrequency contours for doubly periodic medium with unit cell in (a) (green) and for free KL plate (purple). Initial wavevector, $\kappa_{i}$ (red) and the negatively refracted wavevector $\kappa_{r}$ (blue) also shown. (c-d) Schematics crystal couplings as used in Fig.~6(a-b).}
	\label{fig:PlC_Iso}
\end{figure}

\end{document}